\documentclass[nofootinbib,amsmath,notitlepage,preprintnumbers,twocolumn]{revtex4-1}
\usepackage{multirow}
\usepackage{amssymb,esvect,amsmath,graphicx,latexsym,amsthm,slashed,eso-pic}

\newcommand{\beq}{\begin{equation}} \newcommand{\eeq}{\end{equation}}
\newcommand{\bea}{\begin{eqnarray}} \newcommand{\eea}{\end{eqnarray}}

\newcommand{\lsim}{\mathrel{\hbox{\rlap{\lower.55ex\hbox{$\sim$}} \kern-.3em \raise.4ex \hbox{$<$}}}}
\newcommand{\gsim}{\mathrel{\hbox{\rlap{\lower.55ex\hbox{$\sim$}} \kern-.3em \raise.4ex \hbox{$>$}}}}

\def\lsim{\mathrel{\raise.3ex\hbox{$<$\kern-.75em\lower1ex\hbox{$\sim$}}}}
\def\gsim{\mathrel{\raise.3ex\hbox{$>$\kern-.75em\lower1ex\hbox{$\sim$}}}}

\newcommand{\be}{\begin{eqnarray}}
\newcommand{\ee}{\end{eqnarray}}

\newcommand{\benum}{\begin{enumerate}}
\newcommand{\eenum}{\end{enumerate}}
\newcommand{\bi}{\begin{itemize}}
\newcommand{\ei}{\end{itemize}}

\begin{document}

\preprint{FERMILAB-PUB-19-178-A}

\title{Superheavy Dark Matter and ANITA's Anomalous Events}

\author{Dan Hooper$^{a,b,c}$}
\thanks{ORCID: http://orcid.org/0000-0001-8837-4127}

\author{Shalma Wegsman$^{d}$}
\thanks{ORCID: http://orcid.org/0000-0002-4378-842X}

\author{Cosmin Deaconu$^{b}$}
\thanks{ORCID: http://orcid.org/0000-0002-4953-6397}

\author{Abigail Vieregg$^{b,d,e}$}
\thanks{ORCID: http://orcid.org/0000-0002-4528-9886}

\affiliation{$^a$Fermi National Accelerator Laboratory, Theoretical Astrophysics Group, Batavia, IL 60510}
\affiliation{$^b$University of Chicago, Kavli Institute for Cosmological Physics, Chicago, IL 60637}
\affiliation{$^c$University of Chicago, Department of Astronomy and Astrophysics, Chicago, IL 60637}
\affiliation{$^d$University of Chicago, Department of Physics, Chicago, IL 60637}

\affiliation{$^e$University of Chicago, Enrico Fermi Institute, Chicago, IL 60637}

\date{\today}

\begin{abstract}

The ANITA experiment, which is designed to detect ultra-high energy neutrinos, has reported the observation of two anomalous events, directed at angles of $27^{\circ}$ and $35^{\circ}$ with respect to the horizontal. At these angles, the Earth is expected to efficiently absorb ultra-high energy neutrinos, making the origin of these events unclear and motivating explanations involving physics beyond the Standard Model. In this study, we consider the possibility that ANITA's anomalous events are the result of Askaryan emission produced by exotic weakly interacting particles scattering elastically with nuclei in the Antarctic ice sheet. Such particles could be produced by superheavy ($\sim 10^{10}-10^{13}$ GeV) dark matter particles decaying in the halo of the Milky Way. Such scenarios can be constrained by existing measurements of the high-latitude gamma-ray background and the ultra-high energy cosmic ray spectrum, along with searches for ultra-high energy neutrinos by IceCube and other neutrino telescopes.

\end{abstract}

\maketitle

\section{introduction}

The ANtarctic Impulsive Transient Antenna (ANITA) experiment consists of a series of balloon payloads designed to search for broadband, impulsive radio emission produced in the interactions of ultra-high energy neutrinos in the Antarctic ice sheet~\cite{Gorham:2008dv}. Searches for cosmic neutrinos in ANITA's energy range are motivated, in part, by the predictions of a potentially observable flux of cosmogenic neutrinos generated in the interactions of ultra-high energy cosmic rays~\cite{Anchordoqui:2007fi,Kotera:2010yn,Ahlers:2012rz,Taylor:2015rla,AlvesBatista:2018zui,Heinze:2015hhp}. To date, ANITA has completed four flights (ANITA-I, -II, -III and -IV), for a total observation time of 115 days, resulting in the strongest constraints on the diffuse neutrino flux above $\sim 3 \times 10^{10}$ GeV~\cite{Gorham:2019guw,Allison:2018cxu,Gorham:2010kv,Gorham:2008yk}. During the ANITA-I and ANITA-III flights, two high-energy events were observed at angles of $27.4^{\circ}\pm0.3^{\circ}$ and $35.0^{\circ} \pm 0.3^{\circ}$ (with respect to the horizontal), respectively~\cite{Gorham:2016zah,Gorham:2018ydl}. This is surprising given that the Earth is predicted to be highly opaque to neutrinos in ANITA's energy range~\cite{Romero-Wolf:2018zxt}. In light of this, a number scenarios involving physics beyond the Standard Model have been proposed to explain the anomalous events~\cite{Collins:2018jpg,Dudas:2018npp}, including those featuring sterile neutrinos~\cite{Cherry:2018rxj,Huang:2018als,Chauhan:2018lnq}, dark matter decaying near the Earth's core~\cite{Anchordoqui:2018ucj}, and exotic long-lived charged particles~\cite{Connolly:2018ewv,Fox:2018syq}. In this paper, we consider the possibility that ANITA's anomalous events are the result of Askaryan emission produced through the elastic scattering of exotic weakly interacting particles, which are produced in the decays of superheavy dark matter particles in the halo of the Milky Way.

\section{An Askaryan Origin of ANITA's Anomalous Events?}

\begin{figure*}
\includegraphics[width=\textwidth]{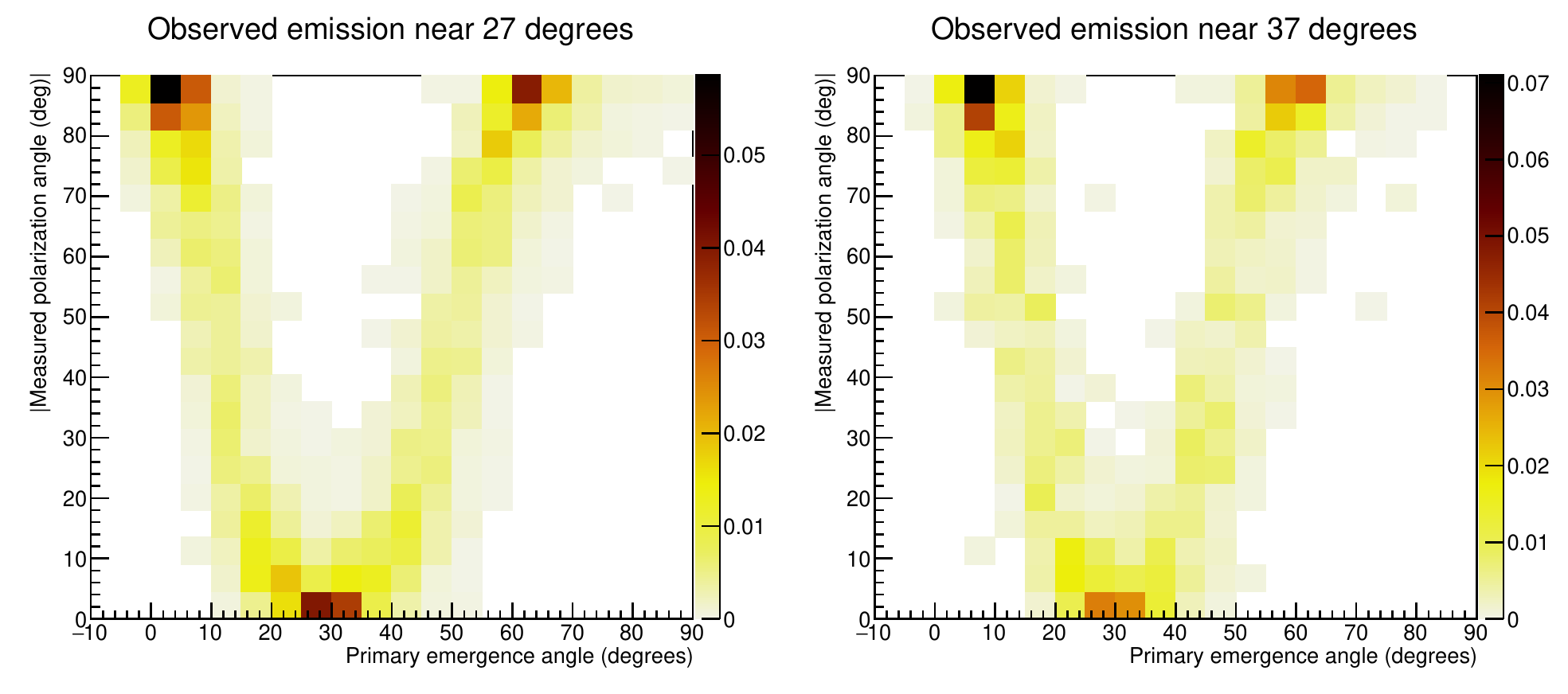}
\caption{The relative acceptance of ANITA-III, as a function of the polarization angle and the emergence angle of the primary, for Askaryan emission observed within 5$^{\circ}$ of the angles of the the two ANITA anomalous events (27$^{\circ}$ and 37$^{\circ}$). Both events are observed to be predominantly horizontally-polarized ($|\theta_{\rm pol}|< 10^\circ$), suggesting that the emission is compatible with primary emergence angles of approximately 30$^{\circ}$.}
\label{fig:polarization}
\end{figure*}

ANITA is capable of observing two classes of events that can arise from upgoing primary particles. The first of these consist of geomagnetic radio emission from extensive air showers, such as those produced by Earth-skimming tau neutrinos. The source of the second class of events is Askaryan emission from showers initiated in the Antarctic ice~\cite{Askaryan:1962hbi}. The radio frequency emission from air shower events is produced as a result of the Earth's magnetic field separating the positively and negatively charged particles, whereas Askaryan emission is the coherent Cherekov light produced at long wavelengths as a result of a negative charge excess in a dense medium.

The ANITA Collaboration has interpreted their two anomalous events as the radio emission from air showers, based on the compatibility of these events with a cosmic-ray air shower waveform shape and their mostly horizontal polarization congruent with the local geomagnetic field. Here, we argue that these events are also compatible with being the Askaryan emission produced by a penetrating particle interacting in the Antarctic ice. In particular, the waveform shape of Askaryan emission~\cite{askaryanIce} is sufficiently similar to that of air showers that the waveform shape alone cannot rule out Askaryan emission for the origin of these events. The observed polarization of Askaryan emission depends on one's location relative to the shower direction. For energetic neutrinos, which can only pass through the Earth at grazing angles due to Earth absorption, the geometry is such that Askaryan emission predominantly produces vertically-polarized events. However, for energetic particles with cross sections that are small enough to enable them to traverse the Earth, Askaryan emission with mostly horizontal polarization is possible. Horizontally polarized Askaryan emission can make waveforms with either polarity, depending on which side of the shower is observed. 

In order to estimate ANITA's acceptance to Askaryan emission from Earth-penetrating particles and determine if an Askaryan interpretation of the two events is possible, we have utilized a modified version of the \texttt{icemc} ANITA Monte Carlo~\cite{icemc}, which includes a full treatment of Askaryan emission from hadronic and electronic showers, propagation of the radio emission to ANITA, and the ANITA instrument response and trigger. \texttt{icemc} is designed to simulate neutrino interactions, but it is possible to disable the effects of neutrino absorption in the Earth in order to approximate the ANITA instrument response to showers from more penetrating primaries.  

\begin{figure*}
    \centering
    \includegraphics[width=0.49\textwidth]{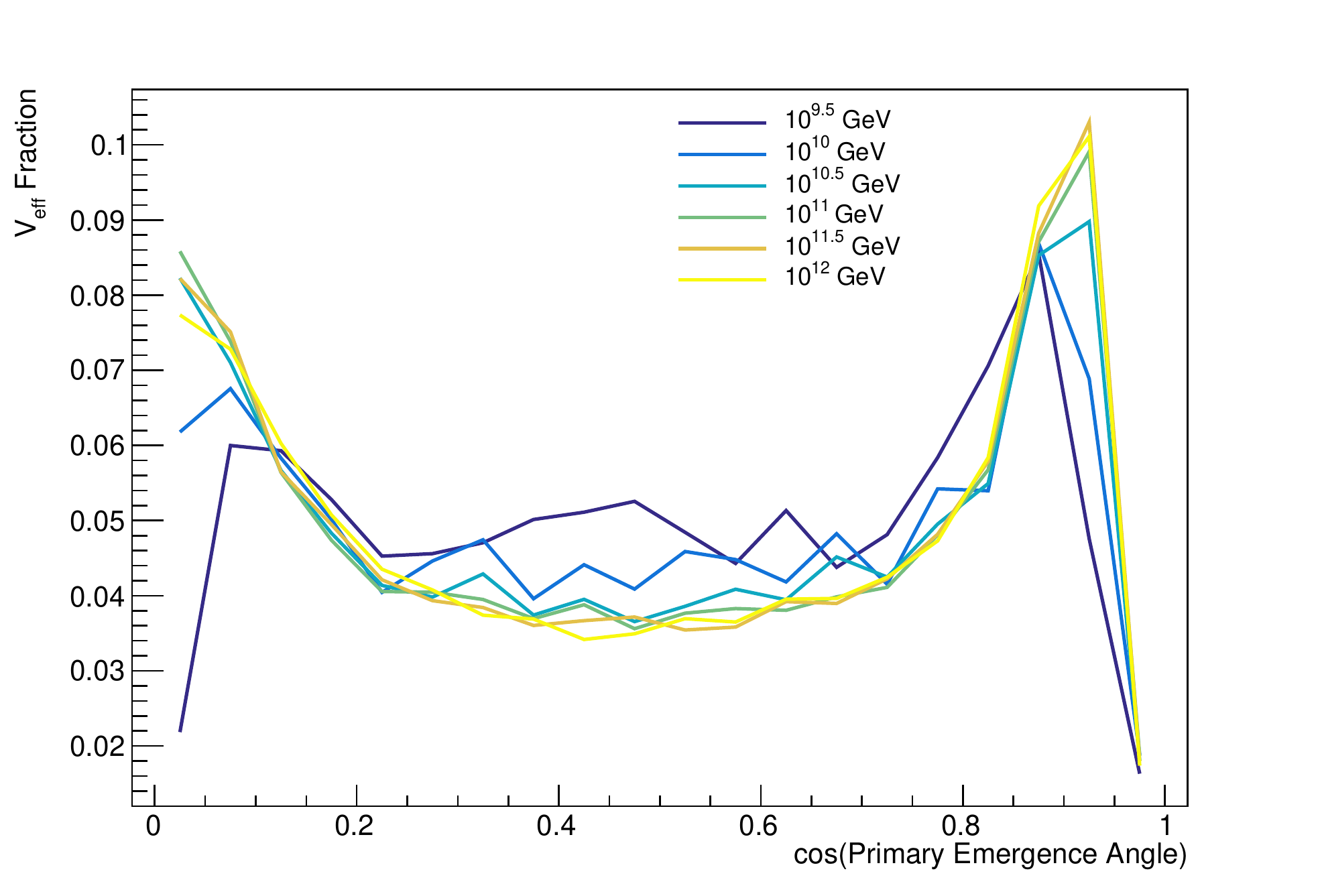}
    \includegraphics[width=0.49\textwidth]{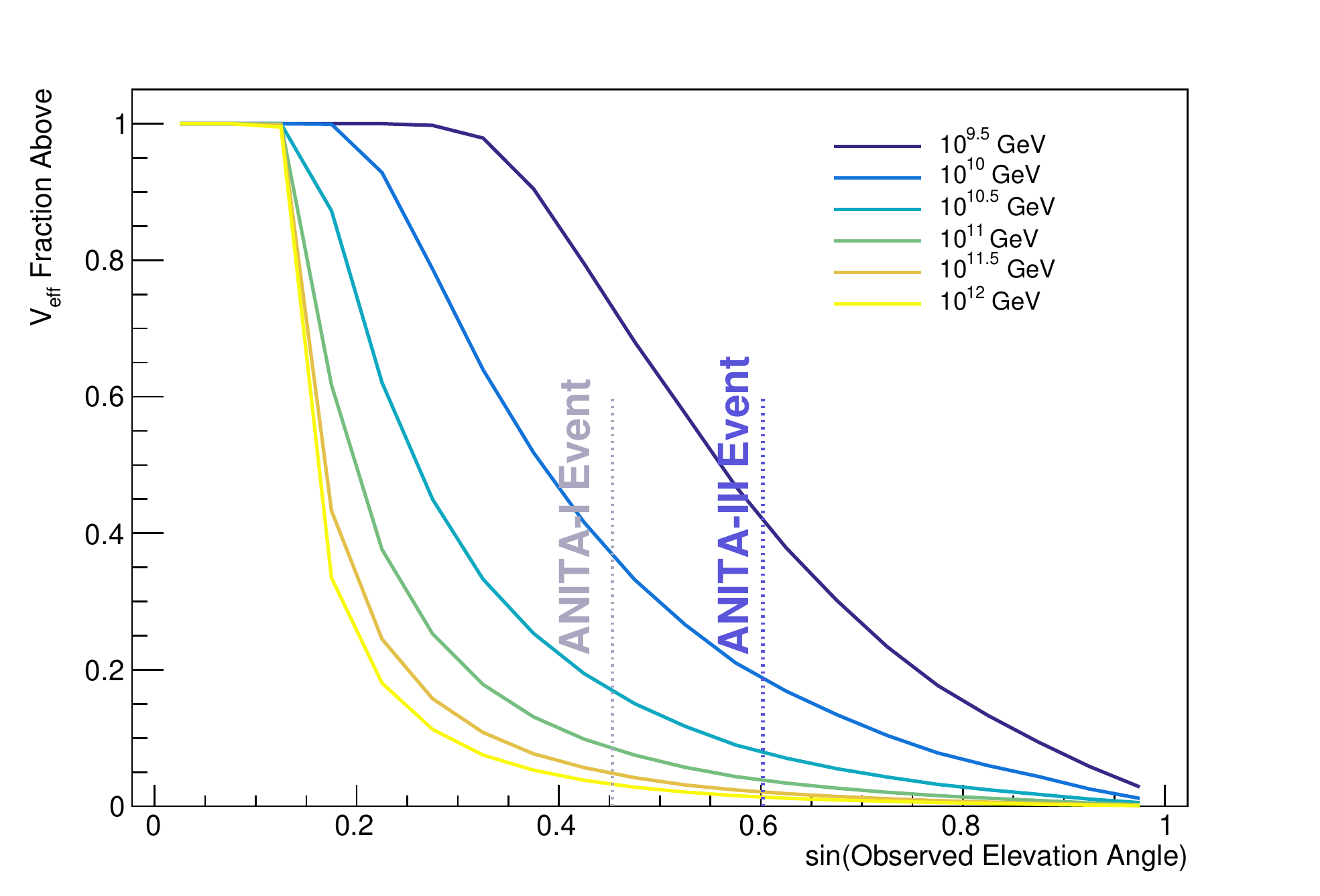}    
    \caption{ANITA-III's relative acceptance as a function of emergence angle (left) and observed elevation angle (right) for a range of shower energies, assuming no earth absorption as is appropriate for a primary particle with small scattering cross section. In the right frame, we mark the elevation angles of the anomalous ANITA-I and ANITA-III events. ANITA's anomalous events can be reasonably interpreted as Askaryan emission from showers with energies in the range of approximately $E_{\rm shower} \sim 10^{9}-10^{11}$ GeV.}
    \label{fig:angles}
\end{figure*}

In Fig.~\ref{fig:polarization}, we plot the relative acceptance of ANITA to showers over a range of energies between $10^9$ and $10^{12}$ GeV as a function of the polarization angle and the emergence angle of the primary particle, for Askaryan emission that is within $5^{\circ}$ of the two anomalous ANITA events ($27^{\circ}$ and $37^{\circ}$, respectively). Combined with the fact that both of these events are mostly horizontally polarized ($|\theta_{\rm pol}|< 10^\circ$), we find that they are each consistent with arising from primaries with emergence angles of approximately $30^{\circ}$. While events with such an emergence angle cannot be produced by ultra-high energy neutrinos, they could be induced by an exotic particle with a smaller interaction cross section. Neglecting absorption in the Earth, ANITA's response is approximately flat with respect to the primary emergence angle, as shown in the left frame of Fig.~\ref{fig:angles}. In the right frame of Fig.~\ref{fig:angles}, we plot ANITA's relative acceptance at different observation angles and for several different shower energies. From these considerations, we conclude that ANITA's anomalous events can be reasonably interpreted as Askaryan emission from showers with energies in the range of approximately $E_{\rm shower} \sim 10^{9}-10^{11}$ GeV, originating from energetic particles which experience relatively little absorption as they propagate through the Earth at an emergence angle of approximately 30$^{\circ}$. We further point out that such an interpretation cannot be excluded by the lack of vertically-polarized Askaryan events observed by ANITA, as each ANITA flight has actually observed such vertically-polarized Askaryan event candidates, although not in significant excess of the background expectation~\cite{Gorham:2019guw,Allison:2018cxu,Gorham:2010kv,Gorham:2008yk}.

\section{Superheavy Dark Matter Decay}

It has long been appreciated that by measuring the angular distribution of high or ultra-high energy cosmic neutrinos, one could use the opacity of the Earth to measure the neutrino-nucleon cross section at energies well beyond those accessible at accelerator experiments~\cite{Kusenko:2002bp,Hooper:2002yq,Borriello:2007cs}. This idea was further exploited in Ref.~\cite{Barbot:2002et}, in which it was proposed that this method could be used to distinguish events produced by exotic weakly interacting particles from those generated by neutrinos. In particular, the authors of Ref.~\cite{Barbot:2002et} considered the detection of ultra-high energy neutralinos produced through the decays of long-lived supermassive dark matter particles in the halo of the Milky Way~\cite{Barbot:2002kh,Berezinsky:1997sb,Ibarra:2002rq,Berezinsky:1997sb}.

In this section, we consider supermassive dark matter particles, $X_d$, that decay to a pair of feebly interacting particles, $\chi$. The energy of each of these particles is in this case simply given by $E_{\chi} = m_{X_d}/2$, and while we do not specify the specific nature of this state we assume that its interaction cross section with nuclei is proportional to that of neutrinos, $\sigma_{\chi N} = f \sigma_{\nu N} \approx f \times 7.8 \times 10^{-36} \, (E/{\rm GeV})^{0.363}$ cm$^2$~\cite{Gandhi:1998ri}. For the time being, we will assume that the $\chi-$nucleon cross section is small enough such that they are not significantly attenuated by the Earth, even at the energies probed by ANITA ($f \lsim 10^{-2}$).

The decays of the $X_d$ population lead to the following flux of ultra-high energy $\chi$'s:
\begin{equation}
F_{\chi}(\Omega) = \frac{2}{4\pi \tau_{X_d} m_{X_d}} \int_{los} \rho_{X_d}(l,\Omega) dl, 
\label{flux1}
\end{equation}
where $\Omega$ is the direction observed, $\tau_{X_d}$ is the lifetime of $X_d$ and the integral is performed over the observed line-of-sight. For the distribution of the $X_d$ population in the halo of the Milky Way, we adopt a Navarro-Frenk-White density profile:
\begin{equation}
\rho_{X_d} \propto \frac{1}{r[1+(r/R_s)]^2},
\end{equation}
where $r$ is the distance to the Galactic Center and we take $R_s=20$ kpc. We normalize the halo such that the local density (at $r=8.25$ kpc) is 0.4 GeV/cm$^3$.

Integrating Eq.~\ref{flux1} over all directions, this scenario yields the following flux (averaged over $4\pi$ sterdians):
\begin{eqnarray}
F_{\chi} \simeq 52\,  {\rm km}^{-2} {\rm yr}^{-1} {\rm sr}^{-1} \times \bigg(\frac{2\times 10^{26} {\rm s}}{\tau_{X_d}}\bigg) \bigg(\frac{10^{11} {\rm GeV}}{m_{X_d}}\bigg). \nonumber \\
\label{flux2}
\end{eqnarray}

\begin{figure}[t]
\includegraphics[scale=0.47]{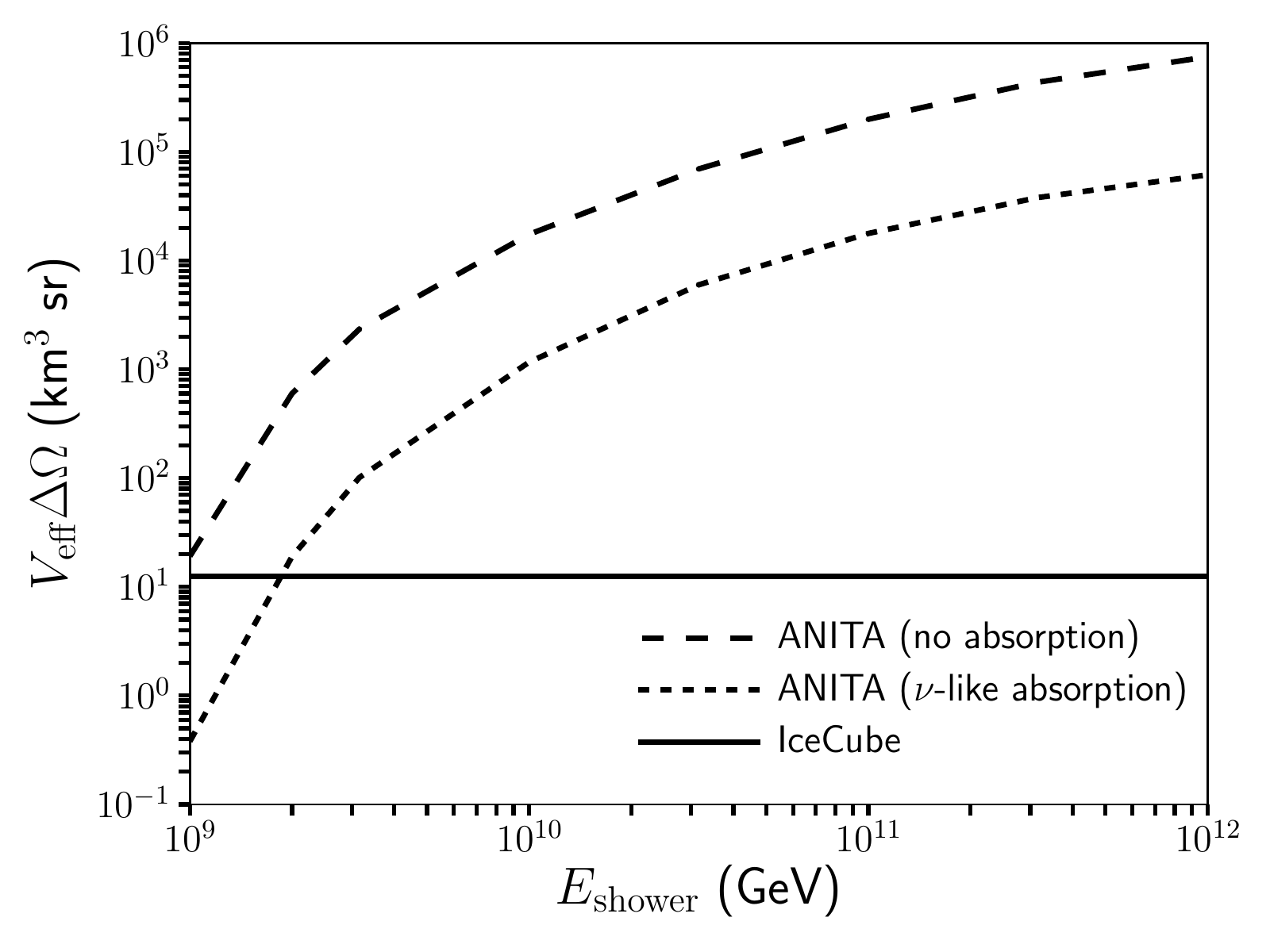} 
\caption{The effective exposure of ANITA and IceCube to ultra-high energy hadronic showers. The dashed line represents ANITA's exposure neglecting any attenuation in the Earth, as is appropriate for showers initiated by particles with very small scattering cross sections, $\sigma_{\chi N} \ll \sigma_{\nu N}$. The dotted curve includes the level of attenuation predicted for Standard Model neutrinos. The effective exposure of IceCube to high-energy showers is approximately 1 km$^3 \times 4\pi$ sr (neglecting any absorption in the Earth). Note that we do not consider IceCube's exposure to muon tracks, as the particles under consideration interact only through elastic scattering.}
\label{effvol}
\end{figure}

In Fig.~\ref{effvol}, we plot the energy-dependent effective exposure for ANITA-III, derived using the \texttt{icemc} ANITA Monte Carlo~\cite{icemc}. The dashed (dotted) curve denotes the exposure to Askaryan events neglecting (including) the effects of neutrino absorption in the Earth. For comparison, we also show the approximate effective exposure of IceCube to high-energy showers, neglecting any absorption in the Earth. Note that we do not consider IceCube's exposure to muon tracks, as the particles in the model under consideration only interact through elastic scattering and thus generate uniquely hadronic shower events.

Combining ANITA's effective exposure with the flux given in Eq.~\ref{flux2}, we can calculate the rate of $\chi$-induced events that will be observed by ANITA:\footnote{In our calculations, we have consider the full 115 days covered by the four ANITA flights. One should keep in mind, however, that the results of ANITA IV's horizontally-polarized channel have not yet been released, and that the ANITA II trigger was less sensitive to horizontally polarized emission than in other flights. The Monte Carlo we have used to estimate ANITA's acceptance to Askaryan showers is based on the performance of ANITA III.}
%
\begin{eqnarray}
N_{\rm events}^{\rm ANITA} &\simeq& F_{\chi} V_{\rm eff} \Delta \Omega \, \sigma_{\chi N}  N_{\rm targets},  \\
&\approx & 2\,   ({\rm per}\, 115 \,{\rm days})  \times \bigg(\frac{4.4 \times 10^{28} {\rm s}}{\tau_{X_d}}\bigg) \nonumber \\
&\times& \bigg(\frac{10^{11} {\rm GeV}}{m_{X_d}}\bigg)^{0.637} \bigg(\frac{V_{\rm eff} \Delta \Omega}{17,430 \, {\rm km}^3 {\rm sr}}\bigg) \bigg(\frac{f}{10^{-2}}\bigg),  \nonumber
\end{eqnarray}
%
where $N_{\rm targets} \simeq 6.0 \times 10^{23}$ cm$^{-3}$ is the number density of nucleons, and in the second line we used the relation $E_{\chi}=m_{X_d}/2$. We have also assumed that $\chi$-nucleon scattering events create a shower of energy $E_{\rm shower} \approx 0.2 E_{\chi}$, similar to that of the neutral-current events of ultra-high energy neutrinos (note that $V_{\rm eff} \Delta \Omega \simeq 17,430$ km$^3$ sr for $E_{\rm shower} = 0.2 \times 0.5 \times 10^{11}$ GeV).

IceCube should also be sensitive to such events, which we estimate would be observed at the following rate:
%
\begin{eqnarray}
N_{\rm events}^{\rm IceCube} &\simeq& F_{\chi} V_{\rm eff} \Delta \Omega\, \sigma_{\chi N}  N_{\rm targets}, \nonumber \\
&\approx & 0.0046\,   {\rm yr}^{-1}  \times \bigg(\frac{4.4 \times 10^{28} {\rm s}}{\tau_{X_d}}\bigg) \bigg(\frac{10^{11} {\rm GeV}}{m_{X_d}}\bigg)^{0.637} \nonumber \\
&\times & \bigg(\frac{V_{\rm eff}}{4\pi \, {\rm km}^3 {\rm sr}}\bigg) \bigg(\frac{f}{10^{-2}}\bigg), \end{eqnarray}
%
where we have adopted an effective exposure for IceCube of $4\pi$ km$^3$ sr.

\begin{figure}[t]
\includegraphics[scale=0.47]{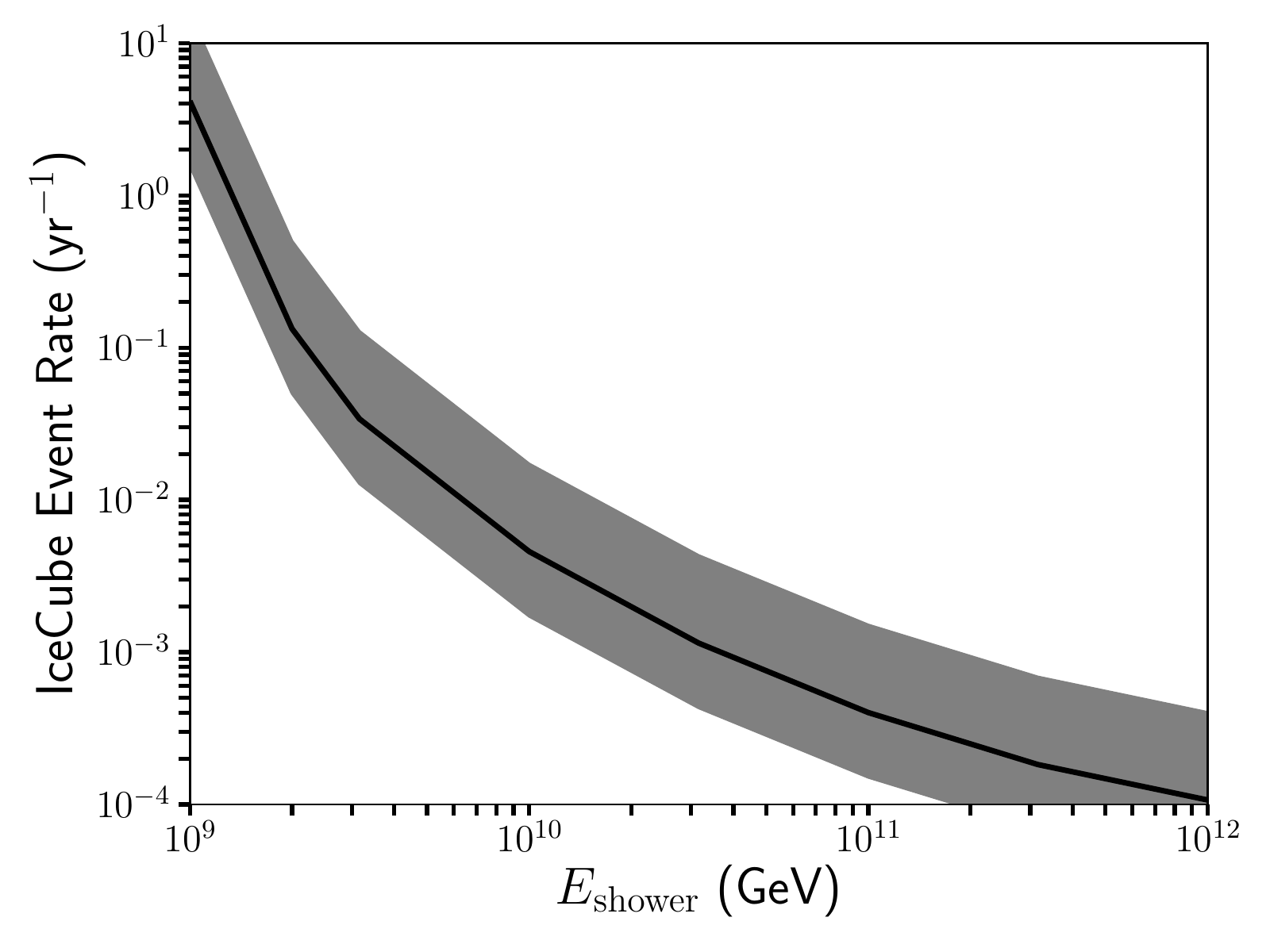} 
\caption{The rate of ultra-high energy shower events at IceCube from the decays of superheavy dark matter into exotic weakly interacting particles, $X_d \rightarrow \chi \chi$, normalizing $f/\tau_{X_d}$ to produce two events over the total flight time of ANITA (115 days). The grey band is the 90\% confidence band around this rate. Given that IceCube has not yet observed any such events, the scenario presented here can explain the two anomalous events observed by ANITA so long as $E_{\rm shower} \gsim (1-2) \times 10^{9}$ GeV, corresponding to $m_{X_d} \gsim (1-2) \times 10^{10}$ GeV.}
\label{icecube}
\end{figure}

In Fig.~\ref{icecube}, we plot the rate of ultra-high energy shower events predicted at IceCube in this scenario, normalizing the value of $f/\tau_{X_d}$ such that ANITA would observe two events over its 115 days of flight time. Given that IceCube has not yet observed any events with an energy greater than $\sim 10^7$ GeV~\cite{Aartsen:2018vtx}, we can constrain this scenario to $m_{X_d} \gsim (1-2) \times 10^{10}$ GeV (and more generally, we can constrain the energy of any Askaryan shower events responsible for ANITA's anomalous events to exceed an energy of $E_{\rm shower} \gsim (1-2) \times 10^{9}$ GeV).

\begin{figure*}[t]
\includegraphics[scale=0.36]{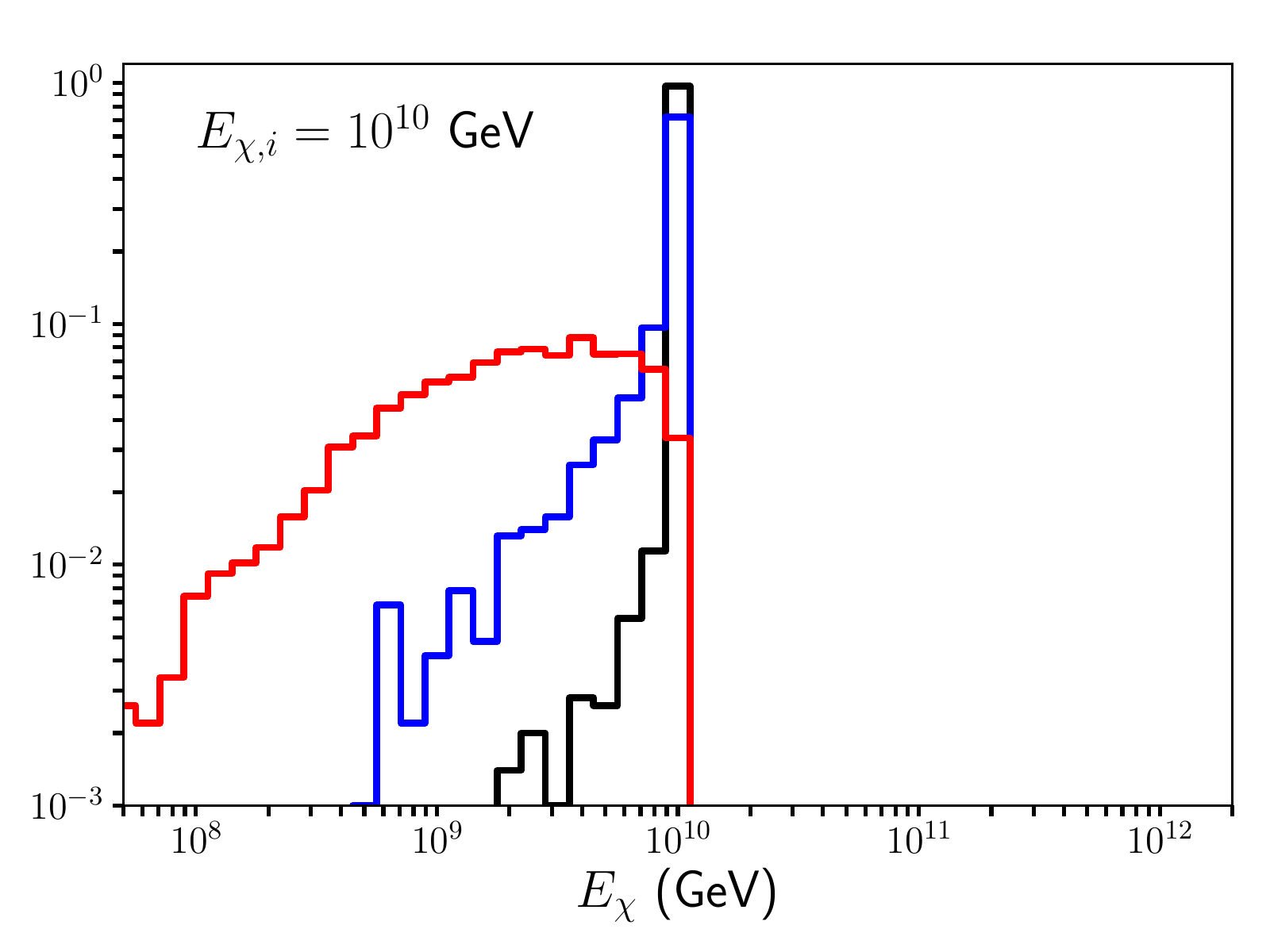} 
\includegraphics[scale=0.36]{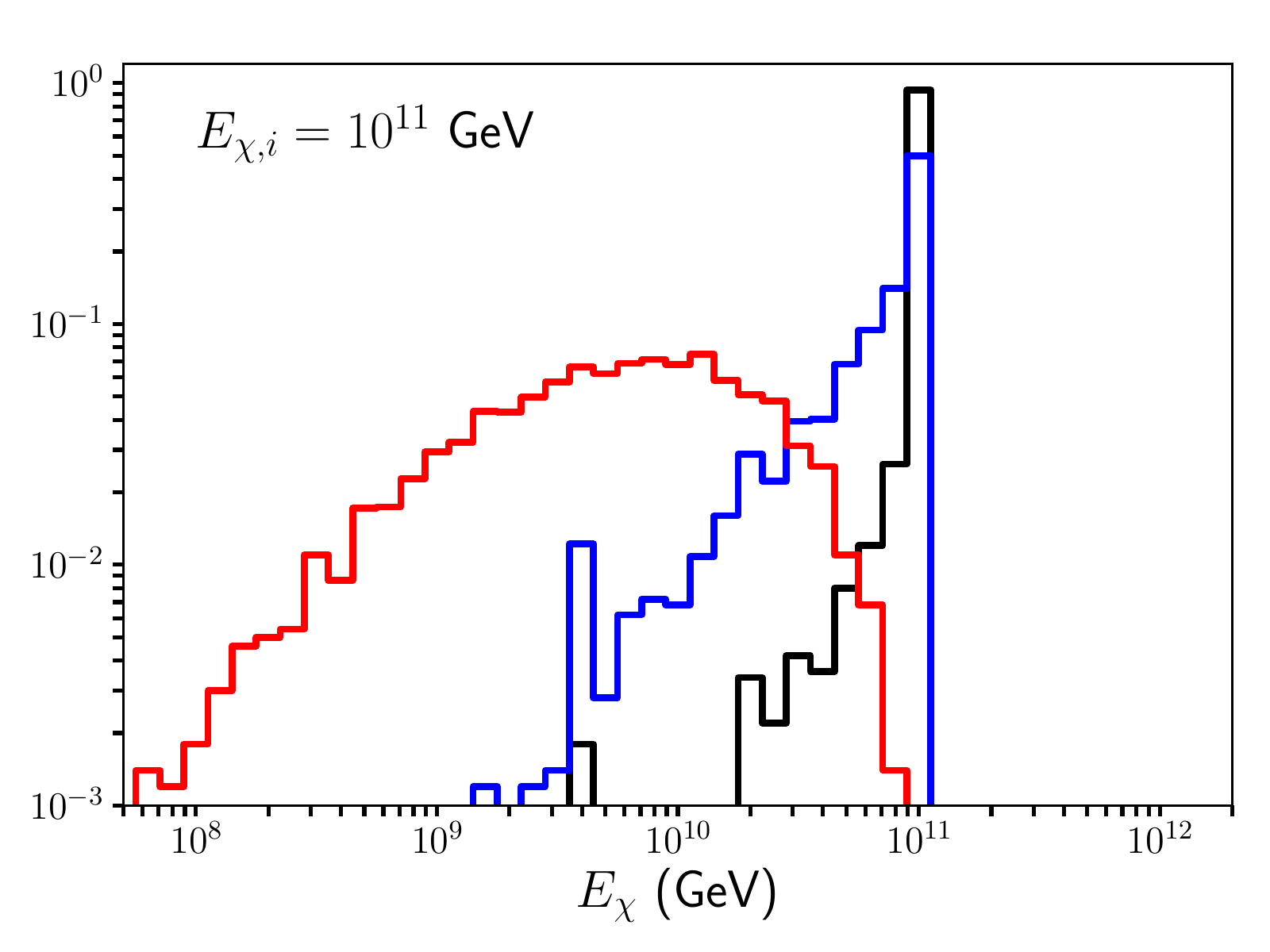} 
\includegraphics[scale=0.36]{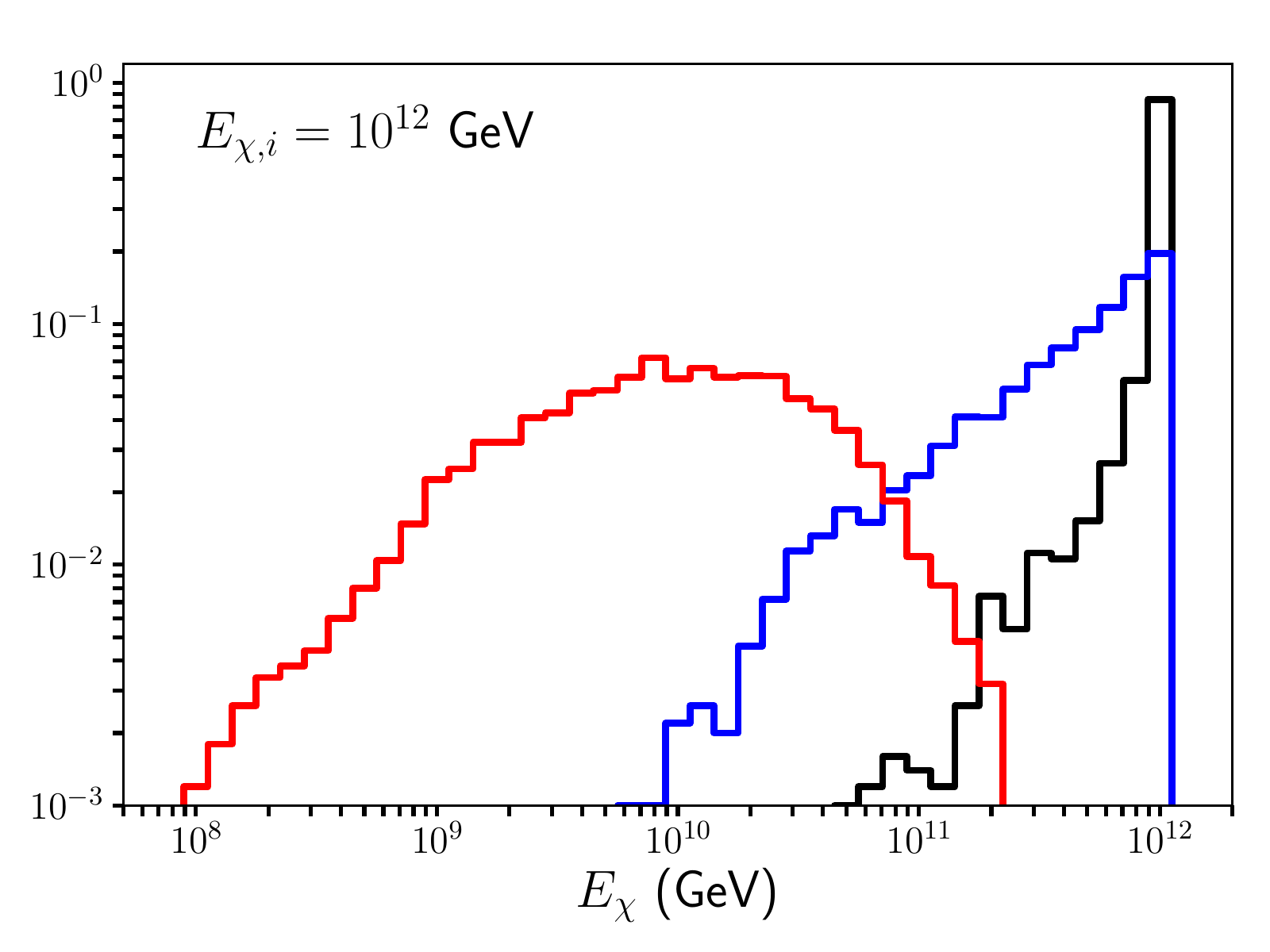} 
\caption{The distribution of $\chi$ energies after passing through the Earth with an emergence angle of $30^{\circ}$. Results are shown for elastic scattering cross sections with nucleons that are 10\% (red), 1\% (blue) or 0.1\% as large as the total neutrino-nucleon cross section ($f=10^{-1}$, $10^{-2}$ or $10^{-3}$). We consider $\chi$'s with an initial energy of $10^{10}$ GeV (left), $10^{11}$ GeV (center) and $10^{12}$ GeV (right).}
\label{histogram}
\end{figure*}  

If the $\chi$-nucleon elastic scattering cross section is not extremely small, such particles may scatter in the Earth, altering the distribution of their energies. In Fig.~\ref{histogram} we plot the $\chi$ energy distribution after passing through the Earth with an emergence angle of $30^{\circ}$, for several values of the initial $\chi$ energy and elastic scattering cross section. For $f \lsim 10^{-2}$, most of these particles do not scatter in the Earth, while the energy distribution is substantially altered in the case of $f\sim 0.1$. For $m_{X_d} \sim \mathcal{O}(10^{10})$ GeV, corresponding to $E_{\rm shower} \sim \mathcal{O}(10^{9})$ GeV, we find that $f$ must be very small in order to avoid tension with the lack of ultra-high energy showers observed by IceCube. For larger values of $m_{X_d}$, larger values of $f$ are possible.

\section{Cosmic-Ray, Gamma-Ray and Neutrino Constraints}

In the previous section, we considered superheavy dark matter that decay uniquely into pairs of exotic weakly interacting particles, $X_d \rightarrow \chi \chi$. It is, of course, possible that such a particle could also decay through other channels to produce Standard Model states. In this section, we consider how cosmic-ray, gamma-ray and cosmic neutrino measurements could be used to constrain the branching fractions of $X_d$ to Standard Model particles (or to particles which decay to Standard Model particles).

The measurement of the high-latitude gamma-ray background by the Fermi Gamma-Ray Space Telescope has been used to place constraints on the lifetime of superheavy dark matter particles, finding $\tau_{X_d} \gsim 10^{28}$ seconds for decays to any combination of Standard Model quarks, charged leptons, or gauge/Higgs bosons~\cite{Blanco:2018esa,Murase:2012xs}. For superheavy dark matter particles that decay into Standard Model states at a rate near this limit, this would also be expected to produce a significant fraction of the highest energy cosmic rays~\cite{Birkel:1998nx,Blasi:2001hr,Protheroe:1996pd,Sigl:1998vz,Bhattacharjee:1991zm,Berezinsky:1997hy,Coriano:2001mg}.

In Ref.~\cite{Barbot:2002kh}, the authors considered a number of examples in which the $X_d$ decays into a combination of quarks, leptons and their superpartners, under the assumption that the low energy particle spectrum is described by the Minimal Supersymmetric Standard Model (MSSM). If we identify the exotic weakly interacting particle in our scenario, $\chi$, as the lightest neutralino~\cite{Bornhauser:2006ve,Anchordoqui:2004qh}, the calculation of the shower evolution~\cite{Barbot:2002ep,Barbot:2002gt,Ibarra:2002rq,Berezinsky:1997sb,Bornhauser:2007sw} yields a spectrum that peaks at $E_{\chi} \sim (0.1-0.2) \, m_{X_d}$ (in $E^2_{\chi} dN_{\chi}/dE_{\chi}$ units). For $m_{X_d}\sim 10^{12}$ GeV, the neutralinos produced in these decays could produce ANITA's two anomalous events for $\tau_{X_d} \sim (0.2-2) \times 10^{28} \, {\rm s} \times (f/0.01)$ (where the precise value depends on the decay channels adopted). In addition to ANITA's anomalous events, such a scenario would also approximately saturate the ultra-high energy cosmic ray spectrum, as well as the constraints based on the observed high-latitude gamma-ray spectrum.

This class of scenarios is most strongly constrained by searches for ultra-high energy cosmic neutrinos, which are predicted to occur at a rate of $\mathcal{O}(10)$ events per year at IceCube~\cite{Barbot:2002kh,Esmaili:2012us,Rott:2014kfa}. The lack of any events at IceCube with an energy greater than $\sim 10^7$ GeV~\cite{Aartsen:2018vtx} limits this class of interpretations. A consistent picture can emerge, however, if one considers a decaying particle with a mass of $\sim 10^{11}-10^{12}$ GeV that decays to exotic weakly interacting particles, possibly along with a small branching fraction ($\sim$$10\%$ percent or less) to Standard Model final states.

\section{Summary and Discussion}

The two anomalous events reported by the ANITA collaboration are puzzling, as neutrinos with enough energy to generate such showers are unable to penetrate the Earth and thus cannot produce events with the observed orientation. A possible way to reconcile the observed features of these events is to consider exotic weakly interacting particles that are capable of traversing the Earth before generating the observed showers. In this paper, we considered the possibility that the anomalous events are the Askaryan emission that is produced through the elastic scattering of exotic weakly interacting particles that are, in turn, produced through the the decays of superheavy dark matter particles in the halo of the Milky Way. Although the measured waveforms and polarizations angles of ANITA's anomalous events are not consistent with Askaryan emission from neutrino induced showers, we demonstrate that they are consistent with Askaryan emission from showers produced in the interactions of exotic weakly interacting particles, to which the Earth is transparent.

We find that superheavy dark matter particles, $X_d$, decaying to exotic weakly interacting particles, $\chi$, could generate ANITA's anomalous events for $m_{X_d} \sim 2 \times 10^{10}-10^{12}$ GeV and a lifetime of $\tau_{X_d \rightarrow \chi \chi} \sim 10^{29}$ s $\times (f/0.01)$, where $f \equiv \sigma_{\chi N}/\sigma_{\nu N}$. If the decays of these superheavy particles also produce Standard Model particles, constraints can be derived from measurements of the high-latitude gamma-ray background and the ultra-high energy cosmic ray spectrum, as well as from the lack of ultra-high energy neutrinos observed by IceCube. Even if the $X_d$ decays do not produce any Standard Model states, IceCube should be able to detect the showers produced through the elastic scattering of $\chi$'s in the Antarctic ice. At present, the lack of ultra-high energy showers observed by IceCube limits this scenario to $m_{X_d} \gsim 2\times 10^{10}$ GeV. Future measurements, such as those by the Gen-2 configuration of IceCube~\cite{Aartsen:2014njl}, will be able to further constrain this class of scenarios.

\begin{acknowledgments}  

Computing resources for this project were provided by the University of Chicago Research Computing Center. This manuscript has been authored by Fermi Research Alliance, LLC under Contract No. DE-AC02-07CH11359 with the U.S. Department of Energy, Office of High Energy Physics.

\end{acknowledgments}

\bibliography{anita}

\end{document}